\UseRawInputEncoding 

\documentclass[journal]{IEEEtran}
\usepackage{SIunitx}
\usepackage{graphicx}
\usepackage{textgreek}

\ifCLASSINFOpdf
\else
\fi
\begin{document}
%
\title{Low-threshold operation of GaAs-based (GaIn)As/Ga(AsSb)/(GaIn)As ``W''-quantum well lasers emitting in the O-band}
%
%
%

\author{Christian~Fuchs, Imad~Limame, Stefan~Reinhard, Jannik~Lehr, J\"org~Hader, Jerome~V.~Moloney, Ada~B\"aumner, Stephan~W.~Koch, Wolfgang~Stolz
\thanks{C.~Fuchs, I.~Limame, S.~Reinhard, J.~Lehr, A.~B\"aumner, S.~W.~Koch, and W.~Stolz are with the Materials Sciences Center and Department of Physics, Philipps-Universität Marburg, Renthof 5, 35032, Marburg, Germany (e-mail: ada.baeumner@physik.uni-marburg.de).}
\thanks{C.~Fuchs and W.~Stolz are with NAsP\textsubscript{III/V} GmbH, Hans-Meerwein-Straße, 35032 Marburg, Germany (e-mail: wolfgang.stolz@nasp.de).}
\thanks{J.~Hader and J.~V.~Moloney are with the College of Optical Sciences, University of Arizona, 1630 E. University Blvd., Tucson, AZ, 85721, USA (e-mail: jhader@acms.arizona.edu).}
\thanks{J.~Hader and J.~V.~Moloney are with Nonlinear Control Strategies Inc., Nonlinear Control Strategies Inc., 7562 N. Palm Circle, Tucson, AZ, 85704, USA.}
\thanks{Manuscript received December XX, 2018; revised January YY, 2019.}}

%
%

\markboth{IEEE PHOTONICS TECHNOLOGY LETTERS, VOL. XX, NO. Y, MONTH ZZ, 20XX}%
{Fuchs \MakeLowercase{\textit{et al.}}: Low-threshold operation of GaAs-based (GaIn)As/Ga(AsSb)/(GaIn)As ``W''-quantum well heterostructure lasers emitting in the O-band}%

%



\maketitle

\begin{abstract}

The influence of the growth conditions as well as the device design on the device performance of (GaIn)As/Ga(AsSb)/(GaIn)As ``W''-quantum well lasers is investigated.
To this purpose, the epitaxy process is scaled to full two inch substrates for improved homogeneity while the growth process is carried out in a single run for an improved quality.
Furthermore, the optical confinement factor is increased by increasing the aluminum concentration within the cladding layers to a value of \SI{65}{\percent}.
The procedure is carried out for devices with emission wavelengths of \SI{1.26}{\micro\metre} as well as \SI{1.30}{\micro\metre}.
Differential efficiencies as high as \SI{58}{\percent} and threshold current densities as low as \SI{0.16}{\kilo\ampere/\centi\square\metre} are observed in case of devices emitting at \SI{1.26}{\micro\metre} at room temperature.
Furthermore, excellent characteristic temperatures of T\textsubscript{0}~=~(72~$\pm$~5)\,\si{\kelvin} and T\textsubscript{1}~=~(293~$\pm$~16)\,\si{\kelvin} are recorded in the temperature range between \SI{10}{\degreeCelsius} and \SI{100}{\degreeCelsius}.
Devices emitting at \SI{1.30}{\micro\metre} exhibit differential efficiencies of \SI{31}{\percent} and threshold current densities of \SI{0.50}{\kilo\ampere/\centi\square\metre} at room temperature.
Further improvements of these properties and wavelength extension schemes are briefly discssused.

\end{abstract}


%
\IEEEpeerreviewmaketitle

\section{Introduction}

\IEEEPARstart{T}{he} development of more efficient semiconductor lasers in the near infrared regime remains a challenging goal due to Auger recombination-based \cite{higashi1999-2} as well as charge carrier leakage-based limitations \cite{chen1983} of present day devices.
Key parameters, which should be optimized in order to reach the above-mentioned goal, are the threshold current density and the temperature stability of novel devices.
The GaAs platform enables the utilization of well-established (AlGa)As/GaAs-based technology.
In these systems, charge carrier leakage can be greatly reduced since the hetero-offsets can be tailored by adapting the aluminum concentration in the respective barrier or cladding materials.
However, dealing with Auger recombination remains challenging as it is defined by intrinsic material properties of the active region itself.
In order to achieve a specific modification of Auger recombination, a change of structural parameters (i.e. compositions and layer thicknesses) needs to be carried out in such a way that the particular combination of the strain state, the transition energy, the spin-orbit split-off energy and the effective masses prohibit Auger recombination.
As a result, studies indicate that the performance of present devices emitting at \SI{1550}{\nano\metre} is still substantially deteriorated by Auger recombination \cite{sweeney1998}.

Two possible approaches towards the suppression of Auger recombination are 1) the application of dilute bismides and 2) the application of \mbox{type-II} heterostructures as active materials \cite{broderick2012, zegrya1995, meyer1998}.
In type-II heterostructures, electrons and holes are spatially separated in adjacent materials \cite{peter1995}.
Consequently, charge carrier properties are dominated by different materials.
Furthermore, the wave function overlap between electron and hole states depends on the actual charge carrier density in the active region because of an electric field arising due to the spatial separation.
The electron-hole wave function overlap is typically improved by utilizing so-called ``W''-quantum well heterostructures (``W''-QWH) \cite{chow2001} rather than simple double-quantum well heterostructures.
``W''-QWHs consist of a hole quantum well sandwiched in between two electron quantum wells resulting in a ``W''-shaped conduction band alignment that results in a higher electron-hole wave function overlap.
The quantum well materials chosen for the present work are (GaIn)As in case of the two electron QWs and Ga(AsSb) for the hole QW.
With this choice of materials, the entire ``W''-QWH can be grown pseudomorphically on GaAs substrates.

Even though these heterostructures are promising candidates for efficient active regions, previous devices were demonstrated at wavelengths, which are too short for telecommunication applications \cite{klem2000, fuchs2016, ryu2002}, had a tendency to switch to higher order transitions as they were operated \cite{zvonkov2013} or exhibited relatively large threshold current densities \cite{fuchs2016}.

The present work aims at reducing the threshold current density in the wavelength range between \SI{1.26}{\micro\metre} and \SI{1.30}{\micro\metre} by optimizing the epitaxy process as well as the structural design of the laser structure.
Improvements are achieved by a)~scaling the epitaxy process to full 2\,inch wafers for an improved homogeneity across the sample, b)~growing the devices in a single epitaxy run for improved material qualities and c)~increasing the aluminum concentration in the cladding layers in order to increase the optical confinement factor.
Additionally, the temperature-dependent performance of devices emitting at the short wavelength end of the O-band is compared to devices emitting right in the center of the O-band.

\section{Experimental setup}

The metal organic vapor phase epitaxy (MOVPE)-based growth of the samples is carried out using an AIXTRON AIX 200 GFR (Gas Foil Rotation) reactor system.
Hydrogen is used as carrier gas in order to transport the group-III sources triethylgallium (TEGa), trimethylindium (TMIn), and trimethylaluminum (TMAl), the group-V sources tertiarybutylarsine (TBAs) and triethylantimony (TESb) as well as the dopant sources diethyltellurium (DETe) and tetrabromomethane (CBr\textsubscript{4}) into the reactor system.
The growth process is carried out on full 2\,inch n-GaAs (001) ($\pm$\,\SI{0.1}{\degree}) substrates after applying a TBAs-stabilized bake out procedure.
Afterwards, a \SI{0.2}{\micro\metre} thick n-GaAs buffer is added to the substrates in order to obtain a high-quality growth surface.
This buffer is followed by a \SI{1.5}{\micro\metre} thick n-(AlGa)As cladding layer and a \SI{0.2}{\micro\metre} thick undoped GaAs separate confinement heterostructure layer.
Two (GaIn)As/Ga(AsSb)/(GaIn)As ``W''-QWHs consisting of \SI{4}{\nano\metre} thick (GaIn)As QWs and a \SI{4}{\nano\metre} thick Ga(AsSb) QW are utilized as active regions for the devices.
These ``W''-QWHs are separated by a \SI{20}{\nano\metre} thick GaAs barrier.
The indium concentration is held constant at \SI{25}{\percent} while the initial aluminum and antimony concentrations of \SI{40}{\percent} and \SI{26}{\percent}, respectively, are increased during this study.
The p-side of the device once again consists of a \SI{0.2}{\micro\metre} thick undoped GaAs separate confinement heterostructure layer and a \SI{1.5}{\micro\metre} thick p-(AlGa)As cladding layer, which is p-doped by applying a decreased group-V/group-III gas phase ratio during the epitaxial growth.
Furthermore, a \SI{0.2}{\micro\metre} thick highly doped p-GaAs cap is added by doping a GaAs layer using CBr\textsubscript{4} in order to ensure small contact resistances.
Furthermore, the growth process of the active region is carried out at a temperature of \SI{550}{\degreeCelsius} while the surrounding layers are deposited using a  growth temperature of \SI{625}{\degreeCelsius}.

Afterwards, the structures are processed using a standard procedure in order to obtain gain-guided broad-area edge-emitting lasers.
More details regarding the device structure as well as the device processing was previously outlined in the literature \cite{fuchs2016}.
The present investigation focuses on the investigation of devices with a cavity length of approximately\SI{1000}{\micro\metre} and a contact width of \SI{100}{\micro\metre}.
The device characterization is performed using laser studies carried out under pulsed operation conditions.
All devices are analyzed using a pulse length of \SI{400}{\nano\second} and a duty cycle of \SI{0.4}{\percent}.
The temperature of the copper heatsink is controlled using a Peltier cooler/heater and can be varied in the temperature range between \SI{10}{\degreeCelsius} and \SI{100}{\degreeCelsius}.
The device output is either detected using a large-area germanium detector in order to obtain laser characteristics or an optical spectrum analyzer in order to obtain spectral information about the laser mode.

\section{Results}

The properties of all three samples investigated in the following sections are summarized in Tab.~\ref{tab:samples}.
They are discussed and compared in the following sections starting with devices emitting at \SI{1.26}{\micro\metre} in Sec.~\ref{sec:1260} followed by the investigation of a device emitting at \SI{1.30}{\micro\metre} in Sec.~\ref{sec:1300}.

\begin{table*}[!ht]
\centering
	\caption{Summary of the properties of samples A, B and C including the cladding aluminium concentration c\textsubscript{Al}, cavity length (L), contact width (W), emission wavelength (\textlambda), differential efficiency (\texteta\textsubscript{d}), threshold current density (j\textsubscript{th}) at a heatsink temperature of \SI{20}{\degreeCelsius}, and characteristic temperatures (T\textsubscript{0}/T\textsubscript{1}).}
  	\label{tab:samples}
	\begin{tabular}{llllllllll}
  	\hline \hline 
	Sample & c\textsubscript{Al} (\si{\percent}) & L (\si{\micro\metre})& W (\si{\micro\metre}) & \textlambda~(\si{\nano\metre}) & \texteta\textsubscript{d} (\si{\percent}) & j\textsubscript{th} (\si{\kilo\ampere/\centi\square\metre}) & T\textsubscript{0} (\si{\kelvin}) & T\textsubscript{1} (\si{\kelvin})\\
	\hline
	A & 40 & 995 & 100 & 1255 & 58 & 0.31 & 85 $\pm$ 3 & 256 $\pm$ 9 \\
	B & 65 & 995 & 100 & 1260 & 54 & 0.16 & 72 $\pm$ 5 & 293 $\pm$ 16 \\
	C & 65 & 1010 & 100 & 1297 & 31 & 0.50 & 76 $\pm$ 7 & 84 $\pm$ 9 \\
	\hline\hline
\end{tabular}
\end{table*}

\subsection{Type-II ``W''-quantum well lasers emitting at \SI{1.26}{\micro\metre}}
\label{sec:1260}

\begin{figure}[!ht]
\centering
	\includegraphics[width = 8.0cm]{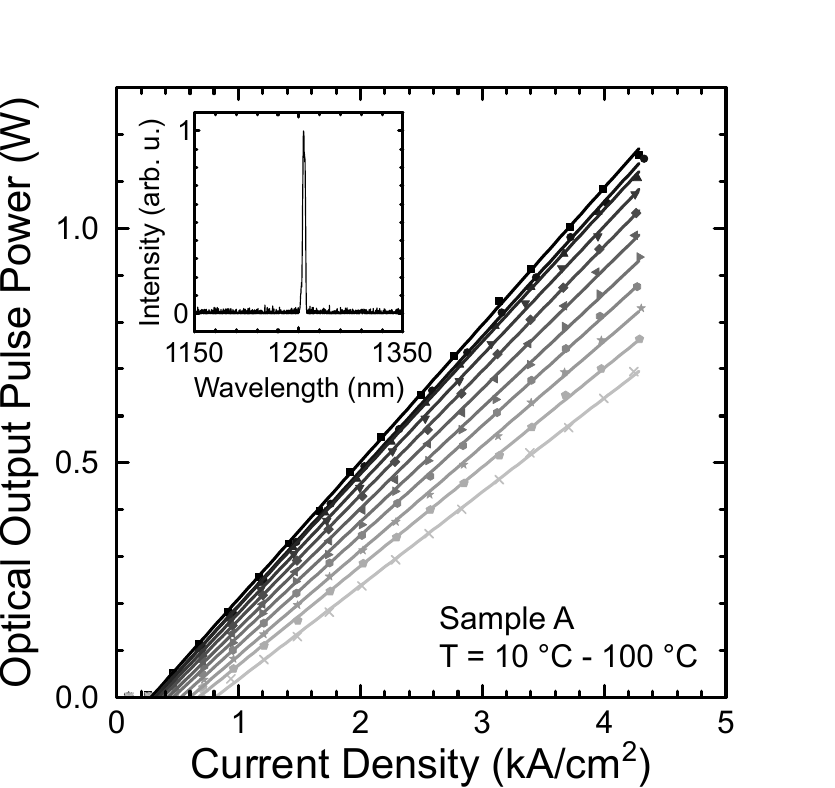}
    \caption{Laser characteristics of sample A for heatsink temperatures between 10 and \SI{100}{\degreeCelsius} in steps of \SI{10}{\degreeCelsius}. A laser characteristic at \SI{20}{\degreeCelsius} was recorded before and after the temperature series in order to verify that no device degradation occurred during the temperature series. The properties obtained from the measurement at \SI{20}{\degreeCelsius} are summarized in Tab.~\ref{tab:samples}. A spectral measurement directly above laser threshold obtained using the OSA system at a temperature of \SI{20}{\degreeCelsius} is given in the inset. This measurement yields an emission wavelength of \SI{1255}{\nano\metre}.}
  \label{fig:Char_1260_Al40}
\end{figure}

Sample A is analyzed by recording laser characteristics in the temperature range between \SI{10}{\degreeCelsius} and \SI{100}{\degreeCelsius} as shown in Fig.~\ref{fig:Char_1260_Al40}.
The measurement at \SI{20}{\degreeCelsius} is carried out before and after the temperature series in order to verify that no device degradation occurs.
Both measurements yielded identical results within the error bar of the experiment and thus, the results obtained from the temperature-dependent data set can be considered as intrinsic device properties.
A differential efficiency of \SI{58}{\percent} and a threshold current density of \SI{0.31}{\kilo\ampere/\centi\square\metre} are deduced from the measurement taken at \SI{20}{\degreeCelsius}.
These results imply an improvement by \SI{+41}{\percent} while the threshold current density is reduced by \SI{-69}{\percent} compared to previous studies \cite{fuchs2018}.
Furthermore, operation based on the fundamental \mbox{type-II} transition is verified by investigating the spectral properties of sample A above threshold in the temperature range between \SI{10}{\degreeCelsius} and \SI{100}{\degreeCelsius}.
The spectrum obtained at a heatsink temperature of \SI{20}{\degreeCelsius} is given as inset in Fig.~\ref{fig:Char_1260_Al40}.
The spectral position of the peak is determined to be \SI{1255}{\nano\metre}.
Consequently, operation based on the fundamental \mbox{type-II} transition is confirmed since these devices were designed for this particular wavelength regime.
These findings indicate that the improved purity and homogeneity by growing these devices in a single epitaxy run on a full 2\,inch wafer in combination with a slightly shorter low-density emission wavelength resulted in substantial improvements of the device characteristics of these \mbox{type-II} lasers.
However, further experiments are required to differentiate between these contributions.

\begin{figure}[!ht]
\centering
	\includegraphics[width = 8.0cm]{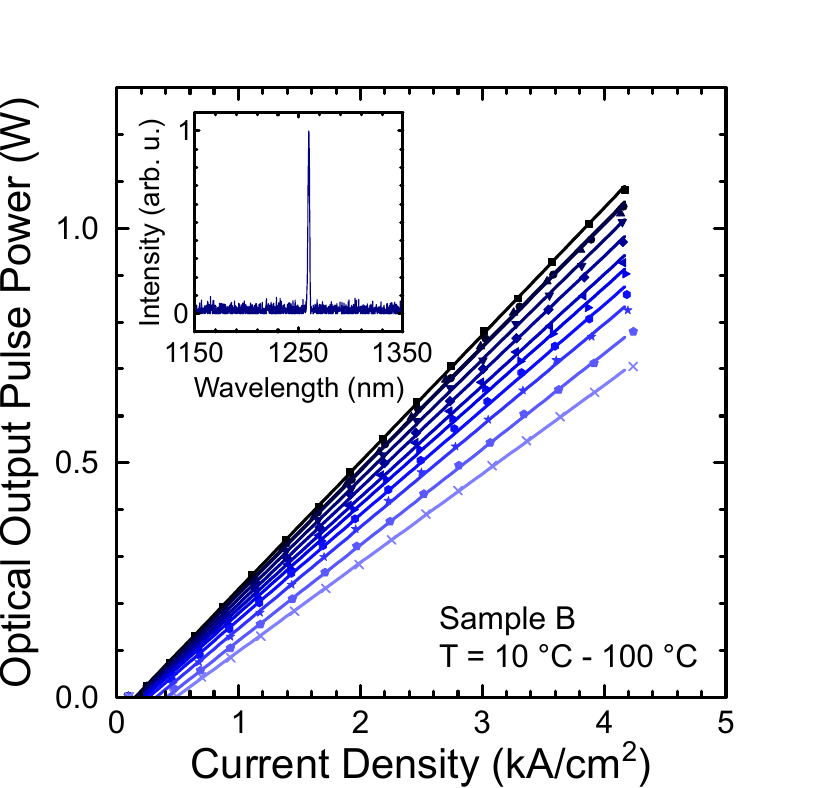}
    \caption{Laser characteristics of sample B for heatsink temperatures between 10 and \SI{100}{\degreeCelsius} in steps of \SI{10}{\degreeCelsius}. A laser characteristic at \SI{20}{\degreeCelsius} was recorded before and after the temperature series in order to verify that no device degradation occurred during the temperature series. The properties obtained from the measurement at \SI{20}{\degreeCelsius} are summarized in Tab.~\ref{tab:samples}. A spectral measurement directly above laser threshold obtained using the OSA system at a temperature of \SI{20}{\degreeCelsius} is given in the inset. This measurement yields an emission wavelength of \SI{1260}{\nano\metre}.}
  \label{fig:Char_1260_Al65}
\end{figure}

Despite these substantial improvements, the above-mentioned threshold current density at \SI{20}{\degreeCelsius} is still higher than the threshold current densities obtained using (GaIn)(NAs) grown on GaAs substrates \cite{tansu2002}.
However, it is important to note that the cavity design of the present devices is different.
The aluminum concentration in the cladding layers is a parameter with a particularly large deviation compared to previous publications resulting in substantially different optical confinement factors.
Consequently, the influence of the aluminum concentration is investigated is investigated by increasing it from \SI{40}{\percent} in sample A to \SI{65}{\percent} in sample B while the active region is grown using the same growth conditions as in case of sample A.
The temperature-dependent measurement of laser characteristics between 10 and \SI{100}{\degreeCelsius} is repeated for sample B and the results are shown in Fig.~\ref{fig:Char_1260_Al65}.
A differential efficiency of \SI{54}{\percent} and a threshold current density of \SI{0.16}{\kilo\ampere/\centi\square\metre} are observed at \SI{20}{\degreeCelsius}.
The increase of the aluminum concentration results in a further reduction of the threshold current density by \SI{48}{\percent} while the differential efficiency remains almost constant with a slight decrease by \SI{-7}{\percent}.
These findings are particularly promising since the reduction of the differential efficiency is comparable to the device-to-device deviations along a single laser bar.
Operation based on the fundamental \mbox{type-II} transition is once again verified for the entire temperature range between \SI{10}{\degreeCelsius} and \SI{100}{\degreeCelsius} and an exemplary spectral measurement of the laser mode at \SI{20}{\degreeCelsius} is shown in the inset of Fig.~\ref{fig:Char_1260_Al65}.
An emission wavelength of \SI{1260}{\nano\metre} is found and thus, low-threshold operation at the short-wavelength end of the O-band is obtained.

\begin{figure}[!ht]
\centering
	\includegraphics[width = 8.0cm]{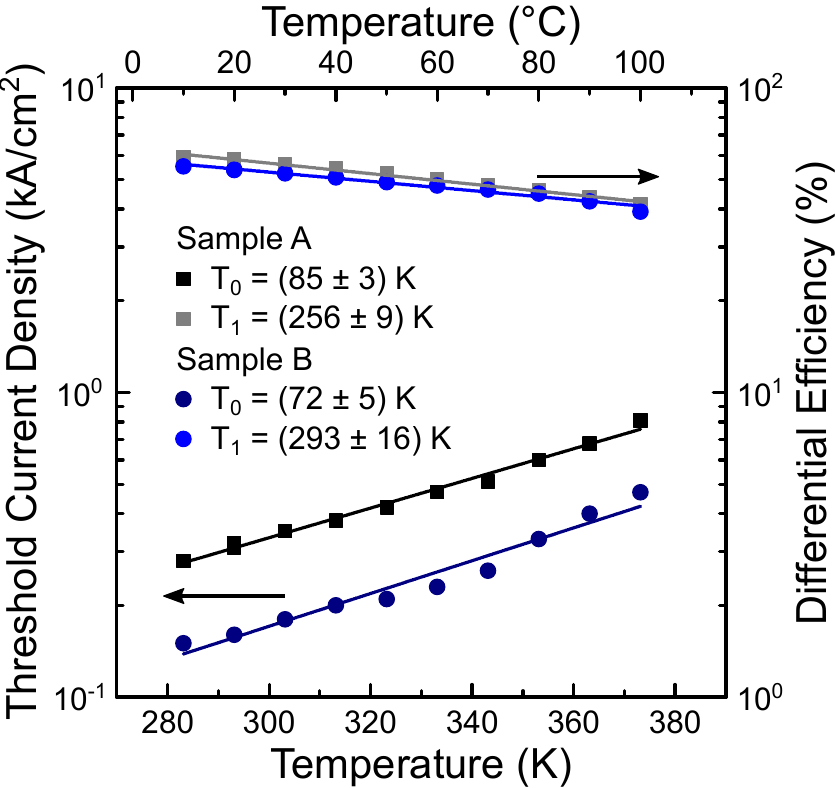}
    \caption{The characteristic temperatures T\textsubscript{0} and T\textsubscript{1} of sample A and B are determined by fitting an exponential model to the temperature-dependent threshold current density and differential efficiency data obtained from the laser characteristics shown in Figs.~\ref{fig:Char_1260_Al40} and~\ref{fig:Char_1260_Al65}. An evaluation yields T\textsubscript{0}~=~(85~$\pm$~3)\,\si{\kelvin} and T\textsubscript{1}~=~(256~$\pm$~9)\,\si{\kelvin} in case of sample A and T\textsubscript{0}~=~(72~$\pm$~5)\,\si{\kelvin} and T\textsubscript{1}~=~(293~$\pm$~16)\,\si{\kelvin} in case of sample B.}
  \label{fig:T0T1_1260}
\end{figure}

The evaluation of the temperature-dependent properties of samples A and B is carried out in the framework of a characteristic temperature model.
Therefore, all laser characteristics shown in Figs.~\ref{fig:Char_1260_Al40} and \ref{fig:Char_1260_Al65} are evaluated and the threshold current density as well as the differential efficiency are plotted as a function of temperature in Fig.~\ref{fig:T0T1_1260}.
The characteristic temperatures T\textsubscript{0} and T\textsubscript{1} describing the temperature-stability of the threshold current density and the differential efficiency, respectively, are obtained applying an exponential model in the entire temperature range.
Characteristic temperatures of T\textsubscript{0}~=~(85~$\pm$~3)\,\si{\kelvin} and T\textsubscript{1}~=~(256~$\pm$~9)\,\si{\kelvin} are found in case of sample A while characteristic temperatures of T\textsubscript{0}~=~(72~$\pm$~5)\,\si{\kelvin} and T\textsubscript{1}~=~(293~$\pm$~16)\,\si{\kelvin} are found in case of sample B.
The differences between the devices are comparably small considering the error bars associated with these characteristic temperatures.
However, the general trend can be expected since lower threshold current densities imply that a given absolute change of the threshold current results in larger relative change resulting in a smaller T\textsubscript{0}.
T\textsubscript{1} is positively affected by low threshold current densities since they result in smaller charge carrier densities in the active region assuming a similar injection efficiency ensuring that the linewidth is not broadened due to band filling among other benefits.
It is also important to note that these single-slope models reproduce the data sets well as opposed to the two-slope model previously used for T\textsubscript{1} \cite{fuchs2018}.
This finding indicates that the decrease of T\textsubscript{1} is driven by the charge carrier density, but a future investigation is still required in order to differentiate between a leakage and a loss process.
Furthermore, a comparison with values obtained using (GaIn)(NAs)/GaAs \cite{tansu2002} yields similar values despite the early stage of development of \mbox{type-II} lasers.

Overall, these results highlight the application potential of \mbox{type-II} lasers due to the excellent threshold current densities, differential efficiencies and characteristic temperatures described above.

\subsection{Type-II ``W''-quantum well lasers emitting at \SI{1.30}{\micro\metre}}
\label{sec:1300}

\begin{figure}[!ht]
\centering
	\includegraphics[width = 8.0cm]{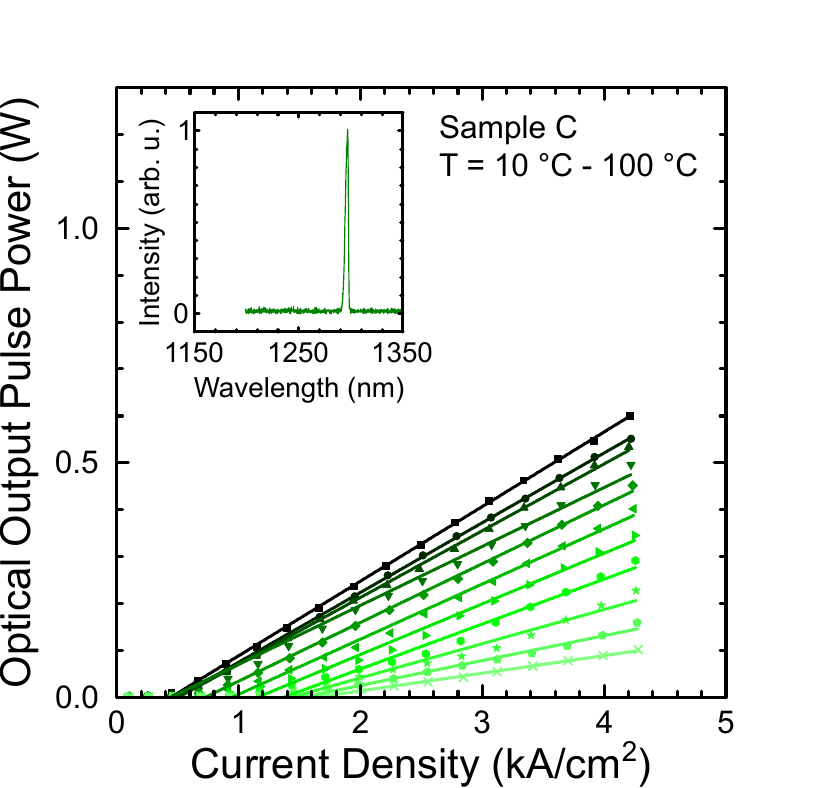}
    \caption{Laser characteristics of sample C for heatsink temperatures between 10 and \SI{100}{\degreeCelsius} in steps of \SI{10}{\degreeCelsius}. A laser characteristic at \SI{20}{\degreeCelsius} was recorded before and after the temperature series in order to verify that no device degradation occurred during the temperature series. The properties obtained from the measurement at \SI{20}{\degreeCelsius} are summarized in Tab.~\ref{tab:samples}. A spectral measurement directly above laser threshold obtained using the OSA system at a temperature of \SI{20}{\degreeCelsius} is given in the inset. This measurement yields an emission wavelength of \SI{1297}{\nano\metre}.}
  \label{fig:Char_1300}
\end{figure}

A further demonstration of devices emitting above \SI{1.26}{\micro\metre} is required in order to eventually cover the entire O-band using such devices.
In order to do so, another device is fabricated by adapting the growth conditions in such a way that the antimony concentration is increased and a red shift of the emission wavelength is obtained.
An aluminum concentration of \SI{65}{\percent} is chosen due to the lower threshold current densities shown above.
The sample is analyzed using the same procedure as the samples outlined above.
Laser characteristics in the temperature range between \SI{10}{\degreeCelsius} and \SI{100}{\degreeCelsius} obtained using sample C are shown in Fig.~\ref{fig:Char_1300}.
The device exhibits a differential efficiency of \SI{31}{\percent} and a threshold current density of \SI{0.50}{\kilo\ampere/\centi\square\metre} at a temperature of \SI{20}{\degreeCelsius}.
A comparison with the devices emitting at \SI{1.26}{\micro\metre} yields lower differential efficiencies and higher threshold current densities.
These results might be caused by the decreased initial wave function overlap at low charge carrier densities, which is reduced due to the more strongly confined holes in the deeper Ga(AsSb) quantum well.
Consequently, a larger carrier density might be required in order to provide sufficient gain for laser operation resulting in deteriorated device properties.
A detailed investigation of this finding using theoretical modeling \cite{berger2015} as well as a systematic variation of the device design is required to understand these findings in more detail.
Furthermore, the laser characteristics are becoming slightly non-linear as the temperature is increased resulting in comparably large error bars of the above mentioned properties.
Spectral measurements at different injection current densities above threshold were carried out and no sign of mode hopping was found in these devices and future investigations are required to understand this behavior in more detail.
A spectral measurement of the laser mode slightly above threshold yields an emission wavelength of \SI{1297}{\nano\metre} as shown in the inset of Fig.~\ref{fig:Char_1300}.
Thus, a successful extension of the emission wavelength towards the center of the O-band is obtained.

\begin{figure}[!ht]
\centering
	\includegraphics[width = 8.0cm]{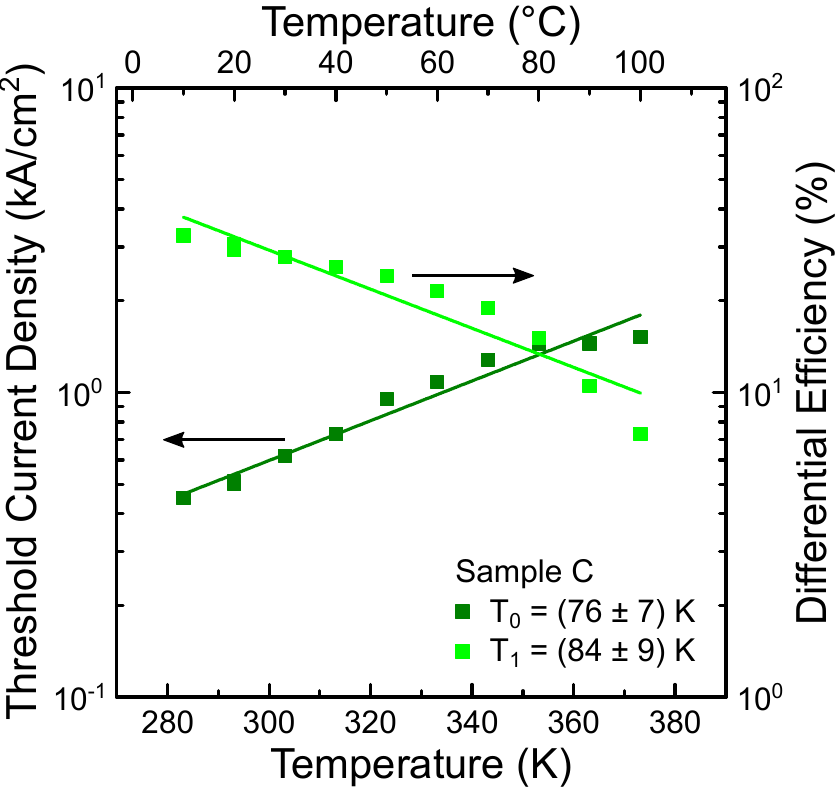}
    \caption{The characteristic temperatures T\textsubscript{0} and T\textsubscript{1} of sample C are determined by fitting an exponential model to the temperature-dependent threshold current density and differential efficiency data obtained from the laser characteristics shown in Figs.~\ref{fig:Char_1300}. An evaluation yields T\textsubscript{0}~=~(76~$\pm$~7)\,\si{\kelvin} and T\textsubscript{1}~=~(84~$\pm$~9)\,\si{\kelvin}.}
  \label{fig:T0T1_1300}
\end{figure}

The reduced device performance at \SI{20}{\degreeCelsius} also translates into less temperature stable devices as shown in Fig.~\ref{fig:T0T1_1300}.
Characteristic temperatures of T\textsubscript{0}~=~(76~$\pm$~7)\,\si{\kelvin} and T\textsubscript{1}~=~(84~$\pm$~9)\,\si{\kelvin} are determined using the same procedure as outlined above.
Furthermore, a non-exponential behavior of T\textsubscript{1} is observed similar to previous studies \cite{fuchs2018}.
A separate analysis of T\textsubscript{1} between \SI{10}{\degreeCelsius} \& \SI{60}{\degreeCelsius} and \SI{70}{\degreeCelsius} \& \SI{100}{\degreeCelsius} yields values of 
T\textsubscript{1}~=~(126~$\pm$~7)\,\si{\kelvin} and T\textsubscript{1}~=~(33~$\pm$~3)\,\si{\kelvin}, respectively.
These results can be explained by the higher threshold current densities and thus, charge carrier densities in the active region, at which the devices are operated.
It is important to note that the errors given above are statistical errors based on the fitting procedure shown in Fig.~\ref{fig:T0T1_1300}.
However, the above-mentioned non-linearities also need to be considered on top of these statistical deviations so that the actual error bar is expected to be larger.
While the operation of this \mbox{type-II} laser based on the fundamental type-II transition up to \SI{100}{\degreeCelsius} is possible, future investigations are required to optimize these heterostructures for emission at \SI{1.30}{\micro\metre} and beyond.
A comparison of the results obtained in Secs.~\ref{sec:1260} and \ref{sec:1300} as well as the literature \cite{fuchs2018} indicate that a substantial improvement of the temperature stability is possible if the threshold current densities are decreased.
Possible strategies towards this goal include a further optimization of the wave guiding structure, the utlization of more active ``W''-QWHs per laser as well as the utilization of different material systems.

\section{Conclusion}

In conclusion, the optimization of the growth conditions of the (GaIn)As/Ga(AsSb)/(GaIn)As type-II ``W''-QWH lasers outlined above yields excellent device properties at an emission wavelength of \SI{1.26}{\micro\metre}.
These properties include a differential efficiency of \SI{58}{\percent} and a threshold current density of \SI{0.31}{\kilo\ampere/\centi\square\metre} at a heatsink temperature of \SI{20}{\degreeCelsius} as well as characteristic temperatures T\textsubscript{0}~=~(85~$\pm$~3)\,\si{\kelvin} and T\textsubscript{1}~=~(256~$\pm$~9)\,\si{\kelvin} determined in the temperature range between \SI{10}{\degreeCelsius} and \SI{100}{\degreeCelsius} based on a standard laser structure \cite{fuchs2018}.
A further improvement is achieved by increasing the aluminum concentration from \SI{40}{\percent} to \SI{65}{\percent} resulting in an increased optical confinement factor.
This change results in a differential efficiency of \SI{54}{\percent} and a threshold current density of \SI{0.16}{\kilo\ampere/\centi\square\metre} at a heatsink temperature of \SI{20}{\degreeCelsius}.
Characteristic temperatures of T\textsubscript{0}~=~(72~$\pm$~5)\,\si{\kelvin} and T\textsubscript{1}~=~(293~$\pm$~16)\,\si{\kelvin} are observed in the above-mentioned temperature range.
Furthermore, an increase of the emission wavelength up to \SI{1.30}{\micro\metre} is demonstrated by increasing the antimony concentration.
At the same time, the device performance at \SI{20}{\degreeCelsius} is decreased as a differential efficiency of \SI{31}{\percent} and a threshold current density of \SI{0.50}{\kilo\ampere/\centi\square\metre} are observed.
This trend is continued in case of the characteristic temperatures, where values of T\textsubscript{0}~=~(76~$\pm$~7)\,\si{\kelvin} and T\textsubscript{1}~=~(84~$\pm$~9)\,\si{\kelvin} are deduced.
However, all devices shown here can be operated based on the fundamental type-II transition up to a temperature of at least \SI{100}{\degreeCelsius}.
Further design changes including the utilization of triple ``W''-QWH lasers, a variation of the separate confinement heterostructure thickness and a fine tuning of the active region will be carried out in order to obtain high-quality devices in the entire O-band.
The progress demonstrated in this publication highlights the application potential of such heterostructures and also paves the way towards the application of novel \mbox{type-II} active regions using different material systems such as dilute nitrides as well as dilute bismides.

\section*{Acknowledgment}

The authors gratefully acknowledge the funding provided by Deutsche Forschungsgemeinschaft (DFG) in the framework of Sonderforschungsbereich 1083 - Structure and Dynamics of Internal Interfaces and the framework of the Research Training Group 1782 -
 Functionalization of Semiconductors.  The Tucson work was supported by the U.S. Air Force Office of Scientific Research under award number \# FA9550-17-1-0246.

\ifCLASSOPTIONcaptionsoff
  \newpage
\fi

\end{document}